\documentclass[aps,prd,twocolumn,superscriptaddress,showpacs,preprintnumbers,amsmath,amssymb]{revtex4}

\usepackage{fancyhdr}
\usepackage{graphicx}
\usepackage{dcolumn}
\usepackage{bm}
\usepackage{longtable}

\begin{document}

\title{Search for anomalous semileptonic decay of heavy flavor hadrons \\ produced in association with a {\boldmath $W$} boson at CDF II}


\date{\today}
\begin{abstract}
We present a search for anomalous semileptonic decays of heavy flavor hadrons produced in association with a $W$ boson, in proton-antiproton collisions at $\sqrt{s}$=1.96$~$TeV. We use $162~{\rm pb}^{-1}$ of data collected with the CDF$~$II detector at the Fermilab Tevatron Collider. We select events with one $W$ boson and at least one jet with an identified secondary vertex.  In the jets with a secondary vertex we look for a semileptonic decay to a muon. We compare the number of jets with both a secondary vertex and a semileptonic decay, and the kinematic properties of these jets, with the standard model expectation of $W$ plus heavy flavor production and decay. No discrepancy is seen between the observation and the expectation, and we set limits on the production cross section of a $B$-like hadron with an anomalously high semileptonic branching ratio.
\end{abstract}
\affiliation{Institute of Physics, Academia Sinica, Taipei, Taiwan 11529, Republic of China} 
\affiliation{Argonne National Laboratory, Argonne, Illinois 60439} 
\affiliation{Institut de Fisica d'Altes Energies, Universitat Autonoma de Barcelona, E-08193, Bellaterra (Barcelona), Spain} 
\affiliation{Baylor University, Waco, Texas  76798} 
\affiliation{Istituto Nazionale di Fisica Nucleare, University of Bologna, I-40127 Bologna, Italy} 
\affiliation{Brandeis University, Waltham, Massachusetts 02254} 
\affiliation{University of California, Davis, Davis, California  95616} 
\affiliation{University of California, Los Angeles, Los Angeles, California  90024} 
\affiliation{University of California, San Diego, La Jolla, California  92093} 
\affiliation{University of California, Santa Barbara, Santa Barbara, California 93106} 
\affiliation{Instituto de Fisica de Cantabria, CSIC-University of Cantabria, 39005 Santander, Spain} 
\affiliation{Carnegie Mellon University, Pittsburgh, PA  15213} 
\affiliation{Enrico Fermi Institute, University of Chicago, Chicago, Illinois 60637} 
\affiliation{Joint Institute for Nuclear Research, RU-141980 Dubna, Russia} 
\affiliation{Duke University, Durham, North Carolina  27708} 
\affiliation{Fermi National Accelerator Laboratory, Batavia, Illinois 60510} 
\affiliation{University of Florida, Gainesville, Florida  32611} 
\affiliation{Laboratori Nazionali di Frascati, Istituto Nazionale di Fisica Nucleare, I-00044 Frascati, Italy} 
\affiliation{University of Geneva, CH-1211 Geneva 4, Switzerland} 
\affiliation{Glasgow University, Glasgow G12 8QQ, United Kingdom} 
\affiliation{Harvard University, Cambridge, Massachusetts 02138} 
\affiliation{Division of High Energy Physics, Department of Physics, University of Helsinki and Helsinki Institute of Physics, FIN-00014, Helsinki, Finland} 
\affiliation{University of Illinois, Urbana, Illinois 61801} 
\affiliation{The Johns Hopkins University, Baltimore, Maryland 21218} 
\affiliation{Institut f\"{u}r Experimentelle Kernphysik, Universit\"{a}t Karlsruhe, 76128 Karlsruhe, Germany} 
\affiliation{High Energy Accelerator Research Organization (KEK), Tsukuba, Ibaraki 305, Japan} 
\affiliation{Center for High Energy Physics: Kyungpook National University, Taegu 702-701; Seoul National University, Seoul 151-742; and SungKyunKwan University, Suwon 440-746; Korea} 
\affiliation{Ernest Orlando Lawrence Berkeley National Laboratory, Berkeley, California 94720} 
\affiliation{University of Liverpool, Liverpool L69 7ZE, United Kingdom} 
\affiliation{University College London, London WC1E 6BT, United Kingdom} 
\affiliation{Massachusetts Institute of Technology, Cambridge, Massachusetts  02139} 
\affiliation{Institute of Particle Physics: McGill University, Montr\'{e}al, Canada H3A~2T8; and University of Toronto, Toronto, Canada M5S~1A7} 
\affiliation{University of Michigan, Ann Arbor, Michigan 48109} 
\affiliation{Michigan State University, East Lansing, Michigan  48824} 
\affiliation{Institution for Theoretical and Experimental Physics, ITEP, Moscow 117259, Russia} 
\affiliation{University of New Mexico, Albuquerque, New Mexico 87131} 
\affiliation{Northwestern University, Evanston, Illinois  60208} 
\affiliation{The Ohio State University, Columbus, Ohio  43210} 
\affiliation{Okayama University, Okayama 700-8530, Japan} 
\affiliation{Osaka City University, Osaka 588, Japan} 
\affiliation{University of Oxford, Oxford OX1 3RH, United Kingdom} 
\affiliation{University of Padova, Istituto Nazionale di Fisica Nucleare, Sezione di Padova-Trento, I-35131 Padova, Italy} 
\affiliation{LPNHE-Universite de Paris 6/IN2P3-CNRS} 
\affiliation{University of Pennsylvania, Philadelphia, Pennsylvania 19104} 
\affiliation{Istituto Nazionale di Fisica Nucleare Pisa, Universities of Pisa, Siena and Scuola Normale Superiore, I-56127 Pisa, Italy} 
\affiliation{University of Pittsburgh, Pittsburgh, Pennsylvania 15260} 
\affiliation{Purdue University, West Lafayette, Indiana 47907} 
\affiliation{University of Rochester, Rochester, New York 14627} 
\affiliation{The Rockefeller University, New York, New York 10021} 
\affiliation{Istituto Nazionale di Fisica Nucleare, Sezione di Roma 1, University of Rome ``La Sapienza," I-00185 Roma, Italy} 
\affiliation{Rutgers University, Piscataway, New Jersey 08855} 
\affiliation{Texas A\&M University, College Station, Texas 77843} 
\affiliation{Istituto Nazionale di Fisica Nucleare, University of Trieste/\ Udine, Italy} 
\affiliation{University of Tsukuba, Tsukuba, Ibaraki 305, Japan} 
\affiliation{Tufts University, Medford, Massachusetts 02155} 
\affiliation{Waseda University, Tokyo 169, Japan} 
\affiliation{Wayne State University, Detroit, Michigan  48201} 
\affiliation{University of Wisconsin, Madison, Wisconsin 53706} 
\affiliation{Yale University, New Haven, Connecticut 06520} 
\author{A.~Abulencia}
\affiliation{University of Illinois, Urbana, Illinois 61801}
\author{D.~Acosta}
\affiliation{University of Florida, Gainesville, Florida  32611}
\author{J.~Adelman}
\affiliation{Enrico Fermi Institute, University of Chicago, Chicago, Illinois 60637}
\author{T.~Affolder}
\affiliation{University of California, Santa Barbara, Santa Barbara, California 93106}
\author{T.~Akimoto}
\affiliation{University of Tsukuba, Tsukuba, Ibaraki 305, Japan}
\author{M.G.~Albrow}
\affiliation{Fermi National Accelerator Laboratory, Batavia, Illinois 60510}
\author{D.~Ambrose}
\affiliation{Fermi National Accelerator Laboratory, Batavia, Illinois 60510}
\author{S.~Amerio}
\affiliation{University of Padova, Istituto Nazionale di Fisica Nucleare, Sezione di Padova-Trento, I-35131 Padova, Italy}
\author{D.~Amidei}
\affiliation{University of Michigan, Ann Arbor, Michigan 48109}
\author{A.~Anastassov}
\affiliation{Rutgers University, Piscataway, New Jersey 08855}
% ...
\collaboration{CDF Collaboration}
\noaffiliation

\maketitle

Anomalous semileptonic decay signatures are predicted by some theories of physics beyond the standard model (SM) \cite{sbquarkth}. A light supersymmetric partner of the bottom quark in a sample of ordinary $b$ quarks could, for instance, enhance the apparent semileptonic branching ratio of $B$ decays. In this scenario, the semileptonic decay of supersymmetric $\tilde{B}$ mesons, characterized by leptons and charmed mesons $\tilde{B}\rightarrow\ell ~\tilde{\nu}~D~X$ or $\tilde{B}\rightarrow \ell ~D~X$ (where $\tilde{\nu}$ indicates a scalar neutrino and $X$ labels possible additional hadrons), could look similar to standard model $B$ decays. In a sample of events with a $W$ boson and at least one jet in Run$~$I of the Tevatron Collider, CDF observed an excess of events in which a single jet contained both an identified secondary vertex and a muon \cite{giromini}. The rate of these events was inconsistent with that expected based on the semileptonic branching fractions of heavy flavor jets ($b$ or $c$). In this analysis we follow up on this observation with the larger data sample of Run$~$II. The CLEO Collaboration has performed a direct search for a supersymmetric $\tilde{B}$ meson with mass between 3.5 and 4.5 GeV/$c^2$, just below the $e^+e^-\rightarrow B\bar{B}$ threshold, finding no evidence for a light scalar bottom quark \cite{cleo}. The D$\O$ Collaboration has also recently reported on a search for excess of semileptonic decay in events with a $W$ boson \cite{d0sj}, setting upper limits on the production of non-SM $Wb\bar{b}$-like and top-like processes.

We present the analysis of a sample of events with associated production of $W$ boson and quarks or gluons (referred to as $W$+ jets in this article). We identify the $W$ boson by its decay to an  isolated electron(muon) carrying large transverse energy(momentum) $E_T$($p_T$) with respect to the beamline, plus a neutrino. We refer to these high $p_T$ electrons or muons as \textquotedblleft primary leptons\textquotedblright. The neutrino escapes the detector and results in an imbalance of total transverse energy, referred to as \textquotedblleft missing $E_T$\textquotedblright ($\not \!\!\!E_T$) \cite{cdfsys}. Quarks produced in association with the $W$ boson hadronize and are observed as jets of charged and neutral particles. We identify the heavy flavor (h.f.) jets from their displaced secondary vertices. Independently, we search for semileptonic decays of h.f. hadrons by looking for a low momentum muon within the jet. At the Tevatron, the primary SM contributions to $W$+h.f. quarks in the final state arise from the production of W$b\bar{b}$, $Wc\bar{c}$, $Wc$ and $t\bar{t}$, with small contributions also from $WZ$ (with $Z\rightarrow b\bar{b},c\bar{c}$) and single top quark production. The dominant background is due to QCD production of multijets, where one of the jets is misidentified as the primary lepton, and there is large $\not \!\!\!E_T $ in the event. Moreover, one of the jets must contain a h.f.\ quark or be misidentified as containing one, referred in this article as \textquotedblleft mistagged\textquotedblright. Smaller background sources arise from $W$ plus a misidentified h.f. jet, $Zb\bar{b}$, $Zc\bar{c}$, $WW$, $ZZ$ and $Z\rightarrow \tau\tau$ events.

The strategy of this analysis can be summarized as follows: We start from the $W$+jets data sample with at least one jet having a displaced secondary vertex. The same sample has been studied and used to measure the $t\bar{t}$ production cross section \cite{secvtx}. In this sample we find the number of events in which at least one jet has both a displaced vertex and a muon, and compare it to expectation. The expectation is estimated from data and Monte Carlo (MC) simulation, determining the probability of finding a muon within the jet with a displaced vertex. This probability depends primarily on the semileptonic decay branching ratio (B.R.) of the $B$, the kinematics of SM processes that lead to $W$+h.f. event, and the efficiency of muon identification. The expectation determined with such an approach is relative to the number of observed events with a displaced secondary vertex, hence we effectively compare the observed rate of semileptonic decay to muons with that expected from the SM processes, and use this comparison to set a limit on $B$-like decays with high semileptonic decay B.R. In addition, sensitivity to new physics with SM-like semileptonic decay B.R., and hence limits on such processes, can be calculated from the comparison of the observed and predicted number of events whose jets have a displaced vertex only, without further requiring a semileptonic decay.

The CDF II detector is described in detail elsewhere \cite{cdf}. The components used in this analysis include the silicon vertex detector (SVX), the central outer tracker (COT), the central electromagnetic and hadronic calorimeter, and the central muon detectors (CMU, CMP and CMX). The data sample, produced in $p\bar{p}$ collisions at $\sqrt{s}$=1.96 TeV  during Run$~$II of the Fermilab Tevatron, was collected by the CDF$~$II detector between March 2002 and August 2003. This analysis is based on an integrated luminosity of 162$\pm$10 pb$^{-1}$ (150$\pm$9$~$pb$^{-1}$ with the CMX detector operational). MC simulations are performed using {\sc PYTHIA v6.2} \cite{phytia} and {\sc HERWIG v6.4} \cite{herwig1,herwig2}, coupled with {\sc CTEQ5L} parton distribution functions \cite{cteq}. Modeling of $b$ and $c$ hadron decay is provided by {\sc QQ v9.1} \cite{qq}. Diboson and $W$+h.f. production are instead modeled with {\sc ALPGEN} \cite{alpgen}, which generates high multiplicity partonic final states using leading order matrix elements, and passed to {\sc HERWIG} for parton showering.

A detailed description of the event selection is given in \cite{secvtx}. Differences specific to this analysis are due mainly to small improvements in the calibration of the muon chambers and the restriction of the primary muon acceptance to the best understood regions of the detector, as done in \cite{SLT}. Candidate primary electrons(muons) from $W$ decay must have $E_T$($p_T$) greater than 20 GeV with $|\eta|<1.1$, and the event must have $\not \!\!\!E_T>20$ GeV, consistent with the presence of a neutrino from the $W$ boson decay. Since leptons from $W$ decay are expected to be isolated from other energy deposits, the isolation ($I$), defined as the calorimeter transverse energy in a cone of $\Delta R\equiv \sqrt{(\Delta\eta)^2+(\Delta\phi)^2}=0.4$ around the lepton (not including the lepton energy itself) divided by the lepton transverse energy/momentum, is required to be less than 0.1. Finally, the $W$+jets dataset is partitioned according to the number of jets with $E_T>15$ GeV and $|\eta|<2.0$ in the event. Jets are identified using a cone algorithm with a cone opening of 0.4 and are constrained to originate from the $p\bar{p}$ collision vertex. The jet energies are corrected to account for variations of the detector response in $\eta$ and time, as well as multiple $p\bar{p}$ interactions.

Heavy quarks ($b$, $c$) are identified by the presence of secondary decay vertices reconstructed with silicon-based tracking information. The search for a secondary vertex is performed jet by jet. An event by event primary vertex nearest to the identified lepton from the $W$ decay is first selected. After identifying a secondary vertex, the two-dimensional decay length $L_{2D}$ \cite{l2d} is calculated. Secondary vertices corresponding to the decay of h.f. hadrons are expected to have large positive $L_{2D}$. A jet with $L_{2D}$ greater than 3 times its uncertainty is said to have a \textquotedblleft SecVtx tag\textquotedblright. The average efficiency for tagging $b$-jets is a function of the jet $E_T$ and ranges from about 20\% for $E_T\simeq 20$ GeV to around 40\% for jets with $E_T$ above 40 GeV \cite{secvtx}.
\begin{table}[ht]
\caption{\small{Total number of $W$+jets events with at least one SecVtx tagged jet, and the estimated sample composition according to the SM contributions. The rightmost column shows the average efficiency of additionally tagging the SecVtx-tagged jet with the SLT.}}
\begin{center}
\begin{ruledtabular}
\begin{tabular}{lccc|c}
$W$+ $n$ jets [$n$]:   &  1 & 2 & $\geq$ 3  & $\epsilon_{\rm SLT}$ [\%] \\
\hline \hline
Total Data  &  157  &  70       & 59   &    - \\
\hline 
$Wb\bar{b}$ &  37$\pm$11  & 19$\pm$6  &  6.0$\pm$1.6  & 7.1$\pm$0.7\\
$Wc\bar{c}$     &  14$\pm$3   & 6.9$\pm$1.9   &  2.0$\pm$0.6  &  4.1$\pm$0.5\\
$Wc$        &  34$\pm$9  &  6.7$\pm$1.7   &  1.5$\pm$0.4  & 4.1$\pm$0.5\\
$Zb\bar{b}$     & 0.7$\pm$0.2   & 0.3$\pm$0.1    & 0.09$\pm$0.09  & 7.1$\pm$0.7 \\
$Zc\bar{c}$ & 0.4$\pm$0.1   & 0.3$\pm$0.1   & 0.09$\pm$0.09 & 4.1$\pm$0.5 \\
$WW$        &  0.6$\pm$0.1  & 1.0$\pm$0.2     &  0.41$\pm$0.06    & 3.1$\pm$0.5\\
$WZ$        &  0.5$\pm$0.1  & 1.0$\pm$0.2     &  0.31$\pm$0.03    & 6.2$\pm$1.3\\
$ZZ$        &   0.0$\pm$0.0 & 0.1$\pm$0.0     &  0.00$\pm$0.00   & 8.8$\pm$2.7\\
$Z\rightarrow\tau\tau$ & 1.1$\pm$0.2 &0.1$\pm$0.1 &  0.00$\pm$0.00   & 3$\pm$4 \\
QCD     & 24$\pm$3  & 9.2$\pm$1.7     & 5.0$\pm$1.1   & 11$\pm$2 \\
$W$+l.f.      &  41$\pm$6 &  15$\pm$2  &  7.1$\pm$0.8 & 1.2$\pm$0.2\\
top ($s$-chann.)  & 0.6$\pm$0.1  & 1.8$\pm$0.2  & 0.7$\pm$0.1 & 7.8$\pm$0.8\\
top ($t$-chann.)  & 2.0$\pm$0.2  & 2.2$\pm$0.3  & 0.6$\pm$0.1 & 7.0$\pm$0.8 \\
$t\bar{t}$  & 0.9$\pm$0.1   & 6.1$\pm$0.5   & 35$\pm$3 & 8.4$\pm$0.8\\
\end{tabular}
\end{ruledtabular}
\end{center}
\label{tab:secvtxcompo}
\end{table}
Table \ref{tab:secvtxcompo} shows the number of events passing our selection requirements with at least one SecVtx-tagged jet. We also show here the expected composition of the observed SecVtx tagged sample as determined by the same techniques used in the $t\bar{t}$ cross section analysis \cite{secvtx}. As in that paper, the $t\bar{t}$ contribution is taken to be the difference between the observed number of events and the total SM expectation, excluding $t\bar{t}$ itself. The corresponding measured $t\bar{t}$ cross section is 5.3$\pm$1.4 pb. The small component from $t\bar{t}$ in the $W$+1,2 jets sample is calculated using the expected ratio of reconstructed $t\bar{t}$ events in $W+\geq 3$ jets, to $W$+1 and $W$+2 jets. Finally, for $W$+1,2 jets, we scale the overall sample composition determined in \cite{secvtx}, and expressed in terms of expected tagged events, in proportion to the observed number of tagged events in this analysis. Hence, the sum of the components in Table \ref{tab:secvtxcompo} is equal to the number of observed events by construction. The scaling between expected and observed events differs from unity by approximately 10\% or less. In Table \ref{tab:secvtxcompo}, QCD refers to non-top QCD multijet production, and $W$+l.f. indicates $W$ plus light flavor jets, where one or more jets has been wrongly tagged by the SecVtx algorithm.

Muons in jets are identified by matching the tracks of the jet, as measured in the COT, with track segments from at least one of the three muon detectors. The distance between the extrapolated track and the muon chamber track segment depends upon muon multiple scattering and, at high $p_T$, intrinsic resolution of the drift chambers. A single quantity ($L$), with the distribution centered around zero and normalized by its expected width, is constructed using parametrization of the matching resolutions directly measured from $J/\psi\rightarrow\mu^+\mu^-$, $Z\rightarrow\mu^+\mu^-$ and $W\rightarrow\mu\nu$ data \cite{SLT}. Candidate tracks are identified as muons if $|L|< 3.5$, where the efficiency for identifying a muon plateaus at about 85\%. We refer to such muons as \textquotedblleft Soft Lepton Tags\textquotedblright or SLT. Backgrounds are preferentially rejected because they are mostly accidental overlaps of uncorrelated tracks and hadrons that  reached the muon chambers, for which the track-stub match is notably worse than for muons. Muons with $p_T>3$ GeV/$c$ are selected and required to be within $\Delta R<0.6$ of a jet axis. The probability of misidentifying a hadron as an SLT muon, denoted as the \textquotedblleft mistag\textquotedblright probability, is measured from data using a large sample of jets in events collected with an inclusive photon trigger ($\gamma$+jets sample). The number of mistags is counted on a per-track basis for jets with $E_T>15$ GeV, and an SLT mistag probability look-up table is constructed as a function of the track $p_T$, $\eta$ and $\phi$. The average mistag probability is approximately 0.7\% per track for tracks above $p_T=3$ GeV/$c$. The tags in the $\gamma$+jets sample include $21.0\pm1.4$\% \cite{SLT} that are due to genuine h.f. decay. We subtract this fraction from all mistag predictions based on the mistag probability look-up table.  The number of tags predicted with this technique agrees to within 10\% \cite{SLT} of the observation in a variety of samples, including QCD multijets and $Z$+jets events. 

We determine the expected number of  events with at least one jet tagged by both SLT and SecVtx (referred to as \textquotedblleft SLT-SecVtx tag\textquotedblright) combining the contributions in Table \ref{tab:secvtxcompo} with the efficiency ($\epsilon_{\rm SLT}$) for finding an SLT muon tag in the same jet tagged by SecVtx. For a h.f. jet, $\epsilon_{\rm SLT}$ depends primarily on the semileptonic B.R., $B\rightarrow \mu~\nu~X$, and the requirement of 3 GeV/$c$ as the muon minimum $p_T$, but it is also affected by the overall efficiency of SLT tagging and correlations with the selected jets. For instance, measured jet energy is underestimated for jets containing muons and neutrinos, and consequently low $E_T$ jets with semileptonic muon decays can fail the $E_T>15$ GeV requirement. This effect is well modeled in the simulation. Since $\epsilon_{\rm SLT}$ depends on kinematic properties of the jets and muons, it is therefore specific to jets from different physics processes. For $Wb\bar{b}$, $Wc\bar{c}$, $t\bar{t}$, single top, diboson events and $Z\rightarrow\tau\tau$, the kinematics are determined from MC simulation and the efficiency for given kinematics from the data. In simulating the SLT tagging, we use both the efficiency of tagging a muon in the jets, measured from data, and the probability of tagging a jet by accidentally tagging a hadron, described by the mistag probability look-up table. The average values of $\epsilon_{\rm SLT}$ are shown in Table \ref{tab:secvtxcompo}. As a cross check of this technique for predicting the number of SLT-SecVtx tags in $W$+jets events, we apply the same procedure to a high purity $b\bar{b}$ sample, derived from SecVtx-tagged dijets and whose composition is measured to be $\sim95$\% $b\bar{b}$ and $\sim5$\% $c\bar{c}$ \cite{lannon}. The number of observed SLT-SecVtx tags in this sample agrees with the prediction to within 10\%.

The value of $\epsilon_{\rm SLT}$ for the QCD background component is determined directly from data. We select events with non-isolated ($I>0.15$) primary lepton, in order to preferentially select QCD events, and require $\not \!\!\! E_T>20~{\rm GeV}$. In QCD events, high $\not \!\!\! E_T$ often results from a neutrino from a semileptonic decay in a h.f.\ jet, or when a jet is pointing towards one of the non-instrumented regions of the calorimeter. The latter typically correspond also to gaps in the absorber material through which hadrons can escape, creating false signals in the muon chambers. Jets separated from the primary lepton are required to have a SecVtx tag, and we measure $\epsilon_{\rm SLT}$ in these jets to be 11$\pm$2\%. This value is independent of the jet multiplicity, the primary lepton's flavor and isolation, while it increases with higher $\not \!\!\! E_T$. Finally, the probability of tagging with the SLT a light flavor jet that has been wrongly tagged by the SecVtx algorithm is determined using a sample of jets with negative $L_{2D}$ in  $W$+jets MC, using the track multiplicity of the jets and the SLT mistag probability.

The number of $W$+jets events with at least one SLT-SecVtx tag is shown in Table \ref{tab:summary}, together with our expectation from SM processes. There is no excess of observed events with SLT-SecVtx tags. The quoted uncertainty in the SM expectation is the total systematic uncertainty. Since we effectively measure the ratio of events with SLT-SecVtx tagged jets to all SecVtx-tagged jets, systematic uncertainties in the event selection, acceptance and reconstruction efficiency cancel out. Systematic uncertainties considered here include the efficiency for SLT tagging muons inside jets from heavy quark fragmentation ($+1/-8$\%), or from mistags ($\pm 10$\%). The uncertainties also include a contribution from MC modeling of particle tracking efficiency in dense jet environments ($\pm5$\%), and a contribution from the MC modeling of the jet energy calibration ($\pm$2\%), which enters as a correlation between the jet $E_T>15$ GeV requirement and the presence of a muon in the jet. Uncertainty on the SM semileptonic decay B.R. of $B$ mesons ($\pm$4.4\%) is also included.

\begin{figure}[ht]
\begin{center}
\begin{tabular}{cc}
\includegraphics[width=4.3cm]{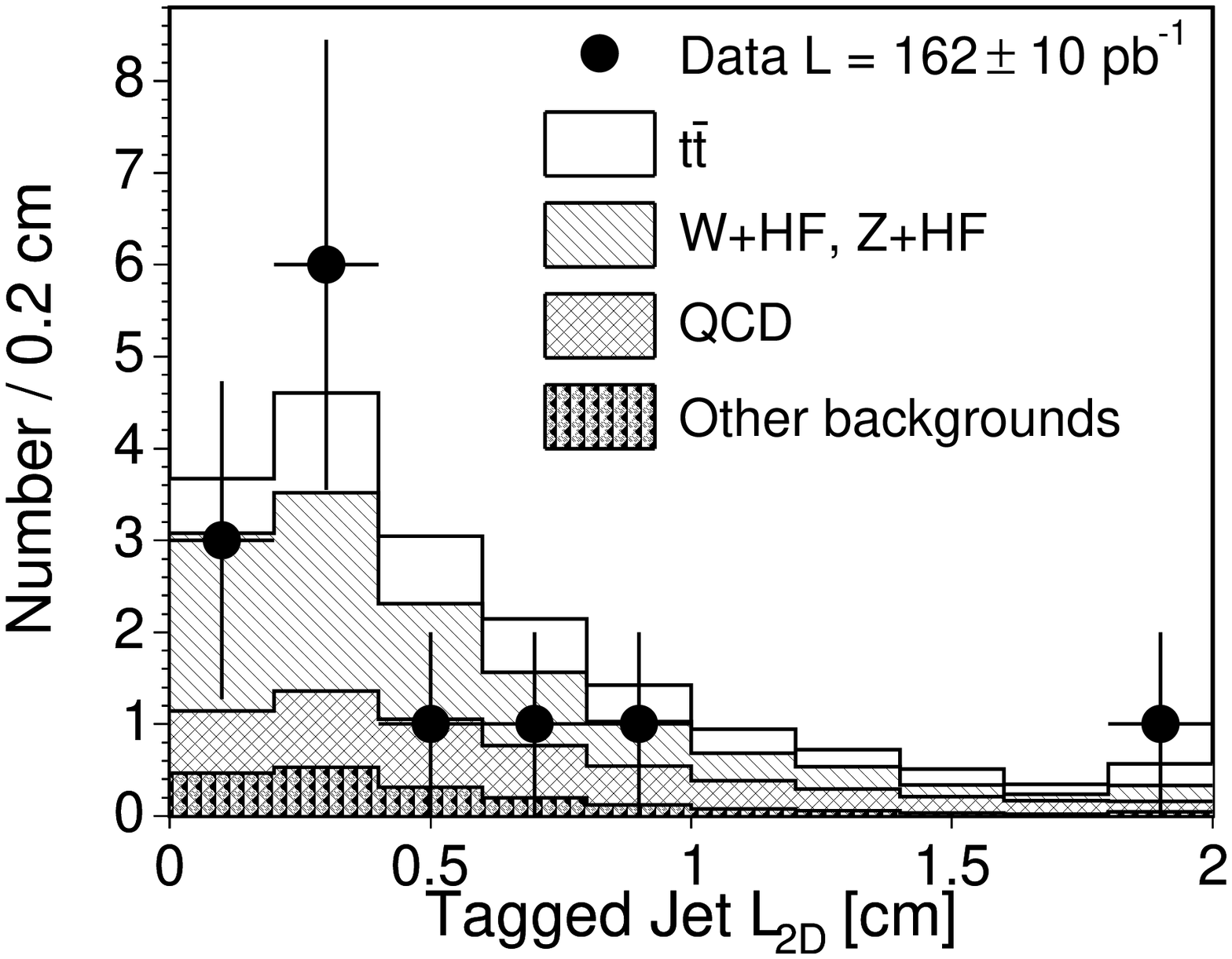} &
\includegraphics[width=4.3cm]{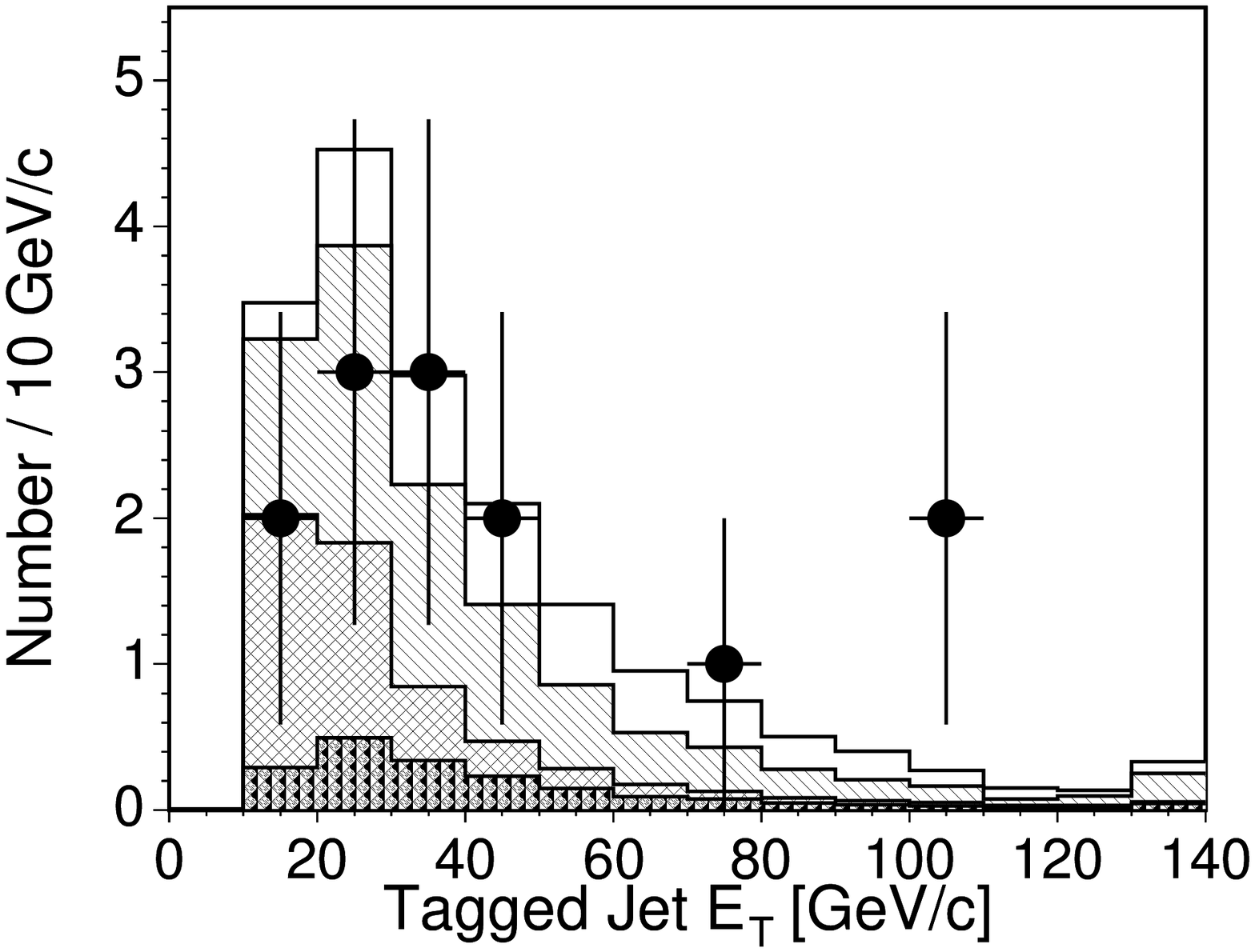} \\
 (a)  & (b) \\
\includegraphics[width=4.3cm]{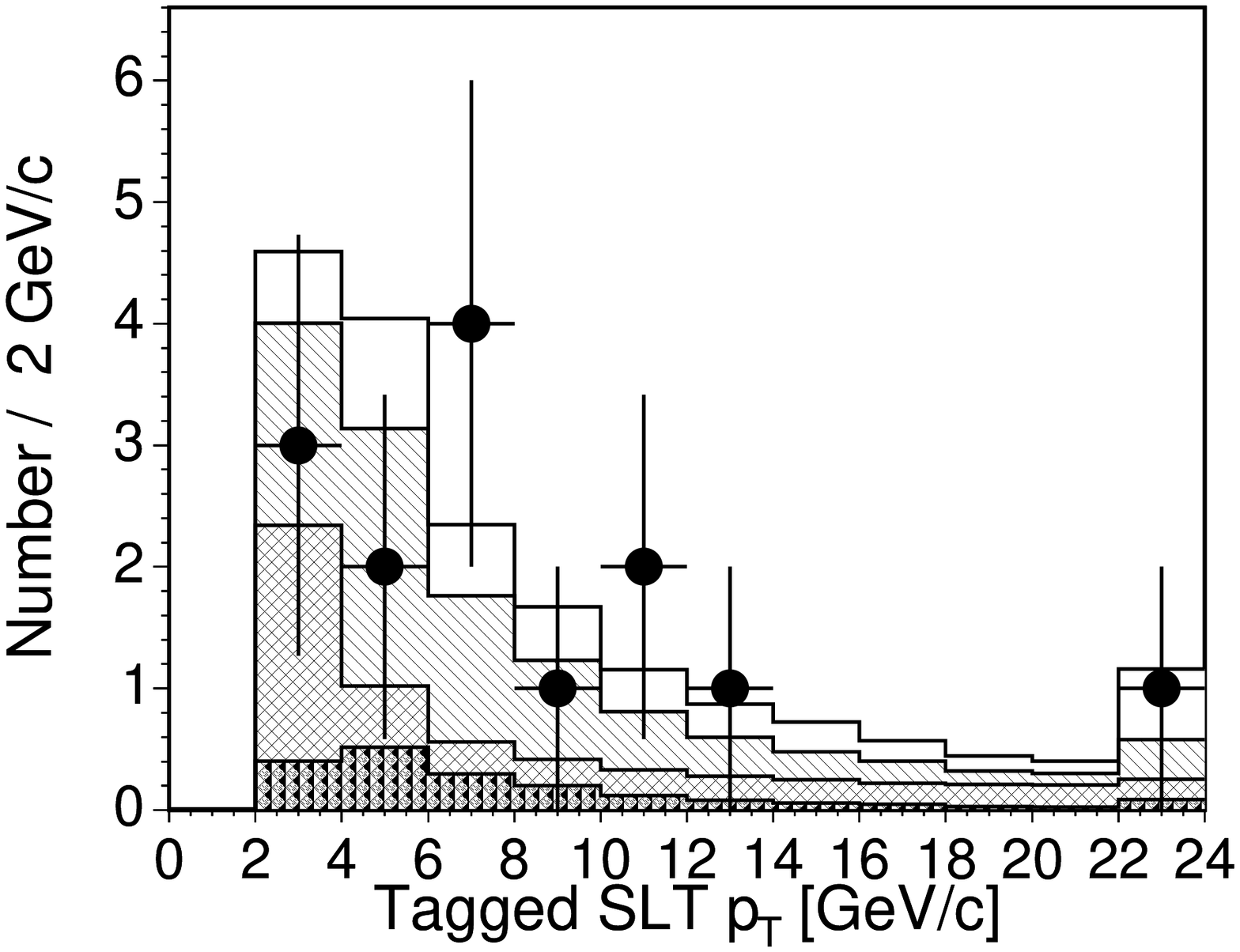} &
\includegraphics[width=4.3cm]{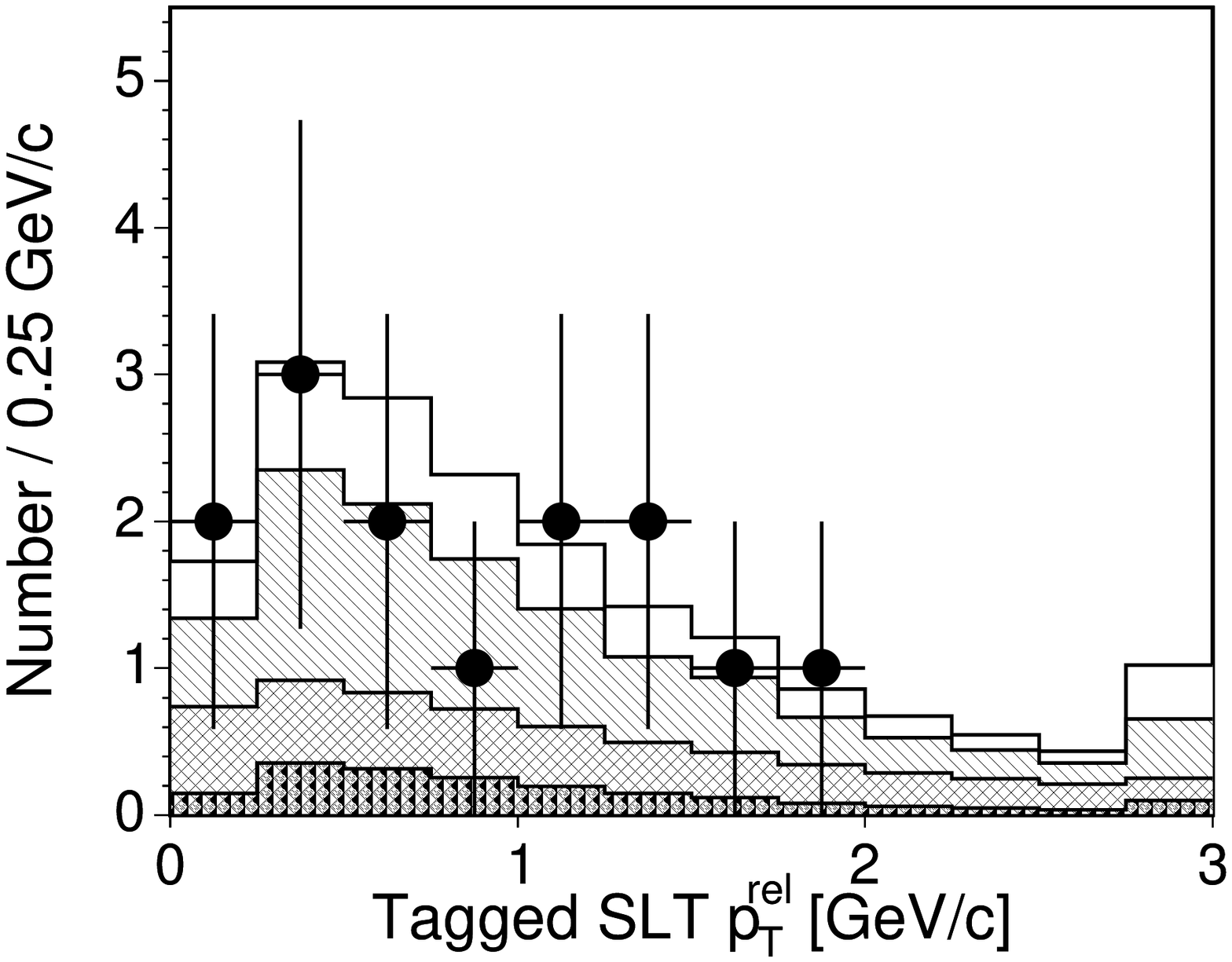} \\
 (c)  & (d) \\
\end{tabular}
\end{center}
\caption{{\small Kinematic distributions for SLT-SecVtx tagged jets.}} 
\label{fig:kinematics}
\end{figure}

We compare the kinematic properties of the SLT-SecVtx tagged sample with SM expectations. Figure \ref{fig:kinematics} shows the distribution of  $L_{2D}$ (a) and $E_T$ (b) for the SLT-SecVtx tagged jets, and $p_T$ (c) and $p_T^{\rm rel}$ (d) for the SLT muons. The quantity $p_T^{\rm rel}$ is the component of the muon momentum orthogonal to the jet axis, which is a discriminant for h.f. decays where the muon can receive a large boost orthogonal to the jet axis. Kolmogorov-Smirnov tests of the shape of the distributions indicate good agreement with the SM predictions. Figures \ref{fig:kinematics}(c) and \ref{fig:kinematics}(d) have one extra data entry because one jet has two identified SLT muons. The shape of the kinematic variable distributions from the SM processes are obtained from MC simulations, with the exception of the QCD component, which is derived from data with the primary lepton isolation $I>0.15$.
\begin{table}[ht]
\caption{\small{Number of expected and observed events with at least one jet tagged by both the SLT and the SecVtx algorithm. The total expected is based on the contributions from all SM processes. The 95\% C.L. limit is calculated for the number of events above SM.}}
\begin{center}
\begin{ruledtabular}
\begin{tabular}{lccc}
$W$+ $n$ jets [$n$]:   &  1  & 2  &  $\geq$3      \\
\hline
$Wb\bar{b}$ & 2.63$\pm$0.80  & 1.56$\pm$0.45  & 0.47$\pm$0.13  \\
$Wc\bar{c}$ & 0.60$\pm$0.15  & 0.27$\pm$0.08  & 0.08$\pm$0.03  \\
$Wc$       &  1.51$\pm$0.40  & 0.26$\pm$0.07  & 0.06$\pm$0.02  \\
Diboson,   &             &  &  \\
$Z\rightarrow\tau\tau$ & 0.09$\pm$0.06 & 0.12$\pm$0.02 & 0.03$\pm$0.01  \\

Z+$b\bar{b}$, $c\bar{c}$ & 0.07$\pm$0.01 & 0.03$\pm$0.01 & 0.01$\pm$0.01  \\
$W$+l.f.    &  0.70$\pm$0.13  & 0.20$\pm$0.04  & 0.08$\pm$0.01  \\
QCD & 2.70$\pm$0.62  & 1.00$\pm$0.26  & 0.55$\pm$0.16  \\
top ($s$+$t$ channel) & 0.19$\pm$0.02 & 0.33$\pm$0.03 & 0.11$\pm$0.02  \\
$t\bar{t}$ & 0.11$\pm$0.01 & 0.69$\pm$0.07 & 3.56$\pm$0.43  \\
\hline
Total Expected & 8.6$\pm$1.2 & 4.5$\pm$0.8 & 4.9$\pm$0.4  \\
Observed & 7 & 5 & 1  \\
Excess 95\% C.L. limit & 6.7         & 6.8             & 3.5    \\
\end{tabular}
\end{ruledtabular}
\end{center}
\label{tab:summary}
\end{table}

Since semileptonic decays in jets produced in association with a $W$ boson are in agreement with the SM rate and kinematics, we proceed to set a limit on non-SM processes that may lead to an excess of SLT muons. We first derive a model-independent limit on the excess of events above SM, as a function of the jet multiplicity. A limit for specific models can be obtained by introducing the model-dependent SLT tagging and event reconstruction efficiency. We derive limits on two kinds of models, one with the signature and kinematics similar to that of SM $Wb\bar{b}$ ($Wb\bar{b}$-like), and one that is similar to top pair production ($t\bar{t}$-like). In each, we introduce a generic model where the non-SM $B$-like object, $\hat{B}$, has all the kinematic properties of a B hadron when it decays semileptonically, but decays with the semileptonic B.R. higher than the SM value. The limits follow from the model-independent case by applying the different SLT tagging and event reconstruction efficiency for the two cases. We set 95\% C.L. limits using a Bayesian approach, with the likelihood given by a Poisson distribution, multiplied by a Gaussian distribution restricted to positive values describing the background, and a constant probability density function representing the signal cross section. Table \ref{tab:summary} gives the model-independent 95\% C.L. limits for the number of $W$+jets events with a SLT-SecVtx tagged jet, coming from anomalous semileptonic decays.

Limits on production cross-sections can be determined by considering the model-specific $\epsilon_{\rm SLT}$, the event reconstruction efficiency and the luminosity of the data sample. Acceptance and efficiency are calculated for 1 and 2 jet events together in the case of a $Wb\bar{b}$-like process, and for $\geq$2 jets in the case of a $t\bar{t}$-like process. As already noted, the efficiency for finding an SLT muon in a SecVtx tagged jet depends on the leptonically decaying species present, their branching ratios and the jet kinematics. In semileptonic $B$ meson decay, a fraction $f_d$ of detected muons originate from the direct decay of the $b$ (with B.R. 10.68$\pm$0.22\% \cite{pdg}), while the rest are from the sequential decay $b\rightarrow c\rightarrow \mu~\nu~X$. The total SLT tagging rate depends on the efficiency ($\epsilon_\mu$) of tagging  the primary and secondary decay muons, with an additional contribution from mistags ($\epsilon_m$). For $Wb\bar{b}$ events we estimate $\epsilon_{\rm SLT}^{Wb\bar{b}}=7.1\pm0.7$\%, with $\epsilon_\mu=5.9\pm0.5$\% and $f_d=56\pm$3\%. For $t\bar{t}$ events we estimate $\epsilon_{\rm SLT}^{t\bar{t}}=8.3\pm0.8$\% with $\epsilon_\mu=6.8\pm0.6$\% and $f_d=58\pm$2\%. 
For a non-SM particle $\hat{B}$ with high semileptonic B.R., the efficiency of tagging with the SLT a SecVtx-tagged jet can therefore also be expressed by: $\epsilon_{\rm SLT}^{\hat{B}} = k\cdot f_d\cdot \epsilon_\mu + \epsilon_m$, where $k$ parametrizes the unknown semileptonic B.R. of the $\hat{B}$, and we assume no contribution from sequential decays. The possibility of multiple SecVtx tagged jets is properly taken into account in the calculation of the efficiency, $\bar{\epsilon}_{\rm SLT}^{\hat{B}}$, for an event with one or more SecVtx tagged jets, to have that same jet additionally SLT tagged. The upper limit on the number of non-SM events present in the SecVtx tagged sample, before requiring SLT tags, is obtained by including $\bar{\epsilon}_{\rm SLT}^{\hat{B}}$ in the Poisson distribution for the signal:
\begin{equation}
P(s|n) = \frac{(s\cdot \bar{\epsilon}_{\rm SLT}^{\hat{B}}+b)^n}{{n!}}e^{-(s\cdot\bar{\epsilon}_{\rm SLT}^{\hat{B}}+b)}, 
\label{equ:likel}
\end{equation}
\noindent where $b$ is the background (i.e. the total SM contribution), $s$ is the non-SM signal before requiring SLT tags, and $n$ is the number of observed SLT-SecVtx tagged events. Since the non-SM events $s$ must be a subset of the observed events shown in Table \ref{tab:secvtxcompo}, the background in the likelihood must also decrease proportionally to the increase of $s$. We introduce the average SM efficiency $\bar{\epsilon}_{\rm SM}=b/N_{\rm tot}$, where $N_{\rm tot}$ is the total number of events with a SecVtx tag but before SLT tagging, so that the numerator of Equation \ref{equ:likel} and the argument of the exponential are replaced by:
\begin{equation}
\begin{tabular}{c}
$s\cdot\bar{\epsilon}_{\rm SLT}^{\hat{B}}+\bar{\epsilon}_{\rm SM}(N_{\rm tot}-s)=s\cdot\left(\bar{\epsilon}_{\rm SLT}^{\hat{B}}-\bar{\epsilon}_{\rm SM}\right)+b.$ \\
\end{tabular}
\label{equ:likel2}
\end{equation}

In Equation \ref{equ:likel2}, the SM contribution decreases as a function of $s$ and the sensitivity of the measurement exists as long as the average efficiency of the non-SM process differs from the efficiency averaged over the SM contributions. This limitation is a direct consequence of the fact that this analysis measures the ratio of SLT-SecVtx tagged over all SecVtx tagged events.

\begin{figure}[t]
\begin{center}
\includegraphics[width=6.5cm]{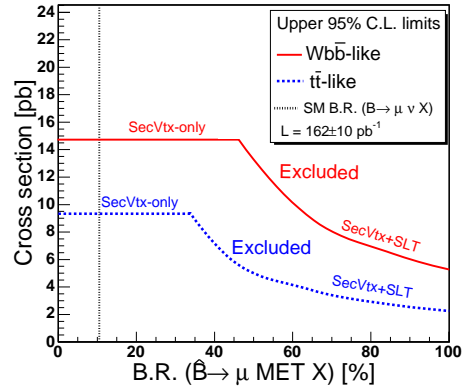}
\end{center}
\caption{\small{Upper limits on the production cross section of W$b\bar{b}$-like and $t\bar{t}$-like processes as a function of the branching ratio: B.R.($\hat{B}$ $\rightarrow \mu~\not \!\!\!E_T~X$).}}
\label{fig:limitssummary}
\end{figure}

Finally, the event efficiency, including the geometric acceptance, the set of cuts applied to the primary lepton and to the jets, as well as the efficiency of the SecVtx algorithm to tag h.f. jets, is measured from a combination of data and MC simulations. Measured efficiencies of the high-$p_T$ lepton triggers are also included in the overall event efficiency. Upper limits in the production cross section are calculated as a function of the semileptonic B.R. and shown in Figure \ref{fig:limitssummary}, indicated by curved lines. For semileptonic B.R.$\sim$100\%, the upper 95\% limit is 5.2 pb for a $Wb\bar{b}$-like model and 2.3 pb for a $t\bar{t}$ model. At lower B.R., the growing limits are truncated because a better limit can be derived simply from the bigger SecVtx tagged sample. In that sample, the absolute normalization on the expectation of SM contributions other than $Wb\bar{b}$ and $t\bar{t}$ is fixed by either MC determination or measured from the data. The 95\% C.L. upper limit on SM $Wb\bar{b}$ production cross section (and consequently any non-SM $Wb\bar{b}$-like production) is calculated to be 14.7 pb. Similarly, the upper limit on $t\bar{t}$  production, and any non-SM $t\bar{t}$-like production, is 9.3 pb. These are shown as horizontal lines in Figure \ref{fig:limitssummary}. For comparison, the analysis in \cite{d0sj} derived upper limits of 26 pb(14 pb) for a $Wb\bar{b}$-like(top-like) process, assuming the SM semileptonic branching fraction. Moreover, based on the excess of events with SLT-SecVtx tags in $W$+2 or more jets, observed during Run I at CDF and considering the differences with the present analysis, we would expect in our dataset an excess of $\sim$12 events due to a non-SM physics process. In $W$+2 or more jets events, we exclude at the 95\% C.L. a non-SM contribution exceeding 5.6 events. 

In summary, we have compared the observed number of $W$+jets events with at least one jet tagged by both the SecVtx and the SLT algorithm, with the expectation based on SM events and the semileptonic B.R. in heavy quark fragmentation. We observe no excess of events above expectation and set a 95\% C.L. upper limit on the production cross section of a hypothetical non-SM particle, with high semileptonic B.R. and kinematic characteristics similar to a $B$ meson in $Wb\bar{b}$ or $t\bar{t}$ events. The limits are set as a function of the semileptonic B.R, providing the most stringent exclusion limits for the production of a non-SM $B$-like meson produced in association with a $W$ boson.

We thank the Fermilab staff and the technical staffs of the participating institutions for their vital contributions. This work was supported by the U.S. Department of Energy and National Science Foundation; the Italian Istituto Nazionale di Fisica Nucleare; the Ministry of Education, Culture, Sports, Science and Technology of Japan; the Natural Sciences and Engineering Research Council of Canada; the National Science Council of the Republic of China; the Swiss National Science Foundation; the A.P. Sloan Foundation; the Bundesministerium f\"ur Bildung und Forschung, Germany; the Korean Science and Engineering Foundation and the Korean Research Foundation; the Particle Physics and Astronomy Research Council and the Royal Society, UK; the Russian Foundation for Basic Research; the Comisi\'on Interministerial de Ciencia y Tecnolog\'{\i}a, Spain; in part by the European Community's Human Potential Programme under contract HPRN-CT-2002-00292; and the Academy of Finland.


\begin{thebibliography}{99.}
\bibitem{cdfsys}We use a ($z,\phi, \theta$) coordinate system where the $z$-axis is in the direction of the proton beam, and $\phi$ and $\theta$ are the azimuthal and polar angles respectively.  The pseudo-rapidity, $\eta$, is defined as $\eta=-\ln(\tan\frac{\theta}{2})$.  The transverse momentum of a charged particle is $p_T = p\sin\theta$, where $p$ represents the measured momentum of the charged-particle track.  The analogous quantity using calorimeter energies, defined as $E_T = E\sin\theta$, is called transverse energy. The missing transverse energy is defined as $\not \!\!\!E_T = \mid-\sum_i E_T^i\hat{n}_i\mid$ where $E_T^i$ is the magnitude of the transverse energy contained in each calorimeter tower $i$ in the pseudo-rapidity region $|\eta|<3.6$, and $\hat{n}_i$ is the direction unit vector of the tower in the plane transverse to the beam direction.
\bibitem{l2d} $L_{2D}$ is the $r$-$\phi$ projection onto the jet axis of the vector pointing from the primary vertex to the secondary vertex. The sign of the projection is chosen relative to the jet direction, i.e. given by ${\rm cos}\phi$, where $\phi$ is the angle between the jet direction and the secondary vertex vector.
\end{thebibliography}
\end{document}